\documentclass[lettersize,journal]{IEEEtran}
\usepackage{amsmath,amsfonts}
\usepackage{algorithmic}
\usepackage{algorithm}
\usepackage{array}
\usepackage[caption=false,font=normalsize,labelfont=sf,textfont=sf]{subfig}
\usepackage{textcomp}
\usepackage{stfloats}
\usepackage{url}
\usepackage{verbatim}
\usepackage{graphicx}
\usepackage{color}
\usepackage{cite}
\begin{document}

\title{Multi-Agent Collaborative Intrusion Detection for Low-Altitude Economy IoT: An LLM-Enhanced Agentic AI Framework}

\author{Hongjuan Li,
        Hui Kang,
        Jiahui Li,
        Geng Sun,~\IEEEmembership{Senior Member,~IEEE,}
        Ruichen Zhang,
        Jiacheng Wang,\\
        Dusit Niyato,~\IEEEmembership{Fellow,~IEEE,}
        Wei Ni,~\IEEEmembership{Fellow,~IEEE,}
        Abbas Jamalipour,~\IEEEmembership{Fellow,~IEEE}
\thanks{
\par Hongjuan Li, Hui Kang, and Jiahui Li are with the College of Computer Science and Technology, Jilin University, Changchun 130012, China, and also with the Key Laboratory of Symbolic Computation and Knowledge Engineering of Ministry of Education, Jilin University, Changchun 130012, China (E-mails: hongjuan23@mails.jlu.edu.cn; kanghui@jlu.edu.cn; lijiahui@jlu.edu.cn).

\par Geng Sun is with the College of Computer Science and Technology, Key Laboratory of Symbolic Computation and Knowledge Engineering of Ministry of Education, Jilin University, Changchun 130012, China, and also with the College of Computing and Data Science, Nanyang Technological University, Singapore 639798 (E-mail: sungeng@jlu.edu.cn).

\par Ruichen Zhang, Jiacheng Wang, and Dusit Niyato are with the College of Computing and Data Science, Nanyang Technological University, Singapore 639798 (E-mails:ruichen.zhang@ntu.edu.sg; jiacheng.wang@ntu.edu.sg; dniyato@ntu.edu.sg).

\par Wei Ni is the School of Engineering, Edith Cowan University, Perth, WA 6027, and the School of Electrical and Data Engineering, the University of Technology Sydney, Sydney, NSW 2007, Australia (E-mail:wei.ni@ieee.org).

\par Abbas Jamalipour is with the School of Electrical and Computer Engineering, The University of Sydney, Sydney, NSW 2006, Australia, and with the Graduate School of Information Sciences, Tohoku University, Sendai 980-8578, Japan (E-mail: a.jamalipour@ieee.org).

\par \textit{(Corresponding authors: Jiahui Li and Geng Sun.)}
}}



\maketitle

\begin{abstract}
The rapid expansion of low-altitude economy Internet of Things (LAE-IoT) networks has created unprecedented security challenges due to dynamic three-dimensional mobility patterns, distributed autonomous operations, and severe resource constraints. Traditional intrusion detection systems designed for static ground-based networks prove inadequate for tackling the unique characteristics of aerial IoT environments, including frequent topology changes, real-time detection requirements, and energy limitations. In this article, we analyze the intrusion detection requirements for LAE-IoT networks, complemented by a comprehensive review of evaluation metrics that cover detection effectiveness, response time, and resource consumption. Then, we investigate transformative potential of agentic artificial intelligence (AI) paradigms and introduce a large language model (LLM)-enabled agentic AI framework for enhancing intrusion detection in LAE-IoT networks. This leads to our proposal of a novel multi-agent collaborative intrusion detection framework that leverages specialized LLM-enhanced agents for intelligent data processing and adaptive classification. Through experimental validation, our framework demonstrates superior performance of over 90\% classification accuracy across multiple benchmark datasets. These results highlight the transformative potential of combining agentic AI principles with LLMs for next-generation LAE-IoT security systems.
\end{abstract}


\section{Introduction}
\label{sec:introduction}
\par Recent advancements in uncrewed aerial vehicle (UAV) and electric vertical take-off and landing (eVTOL) technologies have catalyzed the emergence of the low-altitude economy (LAE), a transformative sector that is reshaping both urban and rural landscapes. At the heart of this evolution lies the LAE Internet of Things (LAE-IoT) network, which enables diverse applications ranging from drone-based package delivery and emergency medical transport to advanced urban air mobility services~\cite{Hamdi2025}. As these applications become increasingly integrated into daily life, ensuring the security and integrity of LAE-IoT networks has evolved from a technical consideration to an essential prerequisite for public safety and societal trust.

\par LAE-IoT networks exhibit distinctive operational characteristics that introduce unprecedented security vulnerabilities. Different from ground-based networks with relatively predictable topologies, aerial networks operate in dynamic three-dimensional (3D) environments with high mobility, distributed autonomous operations, and severe onboard resource constraints~\cite{Mmedhi2025}. This unique combination of factors creates an expanded attack surface that is highly susceptible to sophisticated cyber threats, such as global positioning system (GPS) spoofing that causes aircraft to deviate from designated flight paths or unauthorized altitude changes that compromise mission safety~\cite{Birnbaum2015}. Moreover, adversaries can leverage node mobility and dynamic network composition to deploy malicious nodes during topology transitions, thus enabling coordinated intrusions that evade detection through distributed attack orchestration across aerial nodes~\cite{Ceviz2025}. These sophisticated attack patterns exploit the resource-constrained nature of devices in LAE-IoT networks, where conventional detection systems cannot operate efficiently. This requires adaptive and collaborative detection methods capable of handling rapidly evolving network topologies~\cite{Mmedhi2025}.

\par Traditional intrusion detection systems (IDSs) are ill-suited to tackle the sophisticated security challenges of the low-altitude domain. Specifically, conventional approaches typically rely on static signature-based detection methods and assume stable network topologies, which render them ineffective against the rapidly evolving configurations of aerial swarms and emerging attacks. Moreover, their substantial computational and energy requirements are incompatible with resource-constrained UAVs and other aerial devices, where processing power and battery life are critical operational limitations.

\par To overcome these limitations, this article investigates the transformative potential of large language model (LLM)-enabled agentic artificial intelligence (AI) for developing intelligent, adaptive, and efficient intrusion detection frameworks tailored specifically for LAE-IoT networks. LLMs possess strong cognitive reasoning capabilities that enable a comprehensive understanding of complex cybersecurity patterns, threat taxonomies, and network behaviors~\cite{Gu2025}. When integrated within an agentic framework, these cognitive capabilities empower autonomous agents to conduct proactive threat analysis, adapt dynamically to evolving security landscapes, and collaborate effectively to establish resilient distributed defense systems~\cite{Zisad2025}. This paradigm represents a fundamental shift from reactive, monolithic security approaches toward proactive, collaborative intelligence architectures that hold significant promise for securing the future of LAE operations~\cite{Zambare2025}.

\par Motivated by these considerations, this article explores how LLM-enabled agentic AI can be leveraged to construct a robust intrusion detection framework for LAE-IoT networks. The main contributions are summarized as follows:

\begin{itemize}
    \item We analyze unique challenges of intrusion detection in LAE-IoT networks, including dynamic network topology, real-time detection requirements, and severe resource constraints. This analysis highlights the reasons why traditional methods are ill-suited for these demanding low-altitude aerial environments and establishes the compelling need for a paradigm shift.

    \item We explore how LLM-enabled agentic AI paradigms can overcome the limitations of traditional AI-based approaches in dynamic environments. Moreover, we present a systematic comparison, highlighting proactive, adaptive, and collaborative advantages of agentic AI over reactive, static, and individual detection methods, thus providing strong theoretical justification for our approach.

    \item We propose a general agentic AI-based intrusion detection framework and detail four components of the framework, including perception, memory, reasoning, and action. This agentic AI framework enables proactive and adaptive threat detection, representing a fundamental shift from reactive security approaches to intelligent, self-evolving defense systems.

    \item Through a case study, we present an LLM-enhanced agentic AI system that overcomes the challenges of intrusion detection in LAE-IoT networks by integrating lightweight deployment strategies with LLM-enhanced cognitive capabilities.
\end{itemize}

\section{Overview of LAE-IoT Networks}
\label{sec:overview_laeiot}

\par In this section, we first discuss the key challenges for intrusion detection in LAE-IoT networks. Subsequently, we introduce the evaluation metrics for assessing IDS performance and analyze the limitations of current IDSs.

\begin{figure*}
    \centering
    \includegraphics[width=0.95 \linewidth]{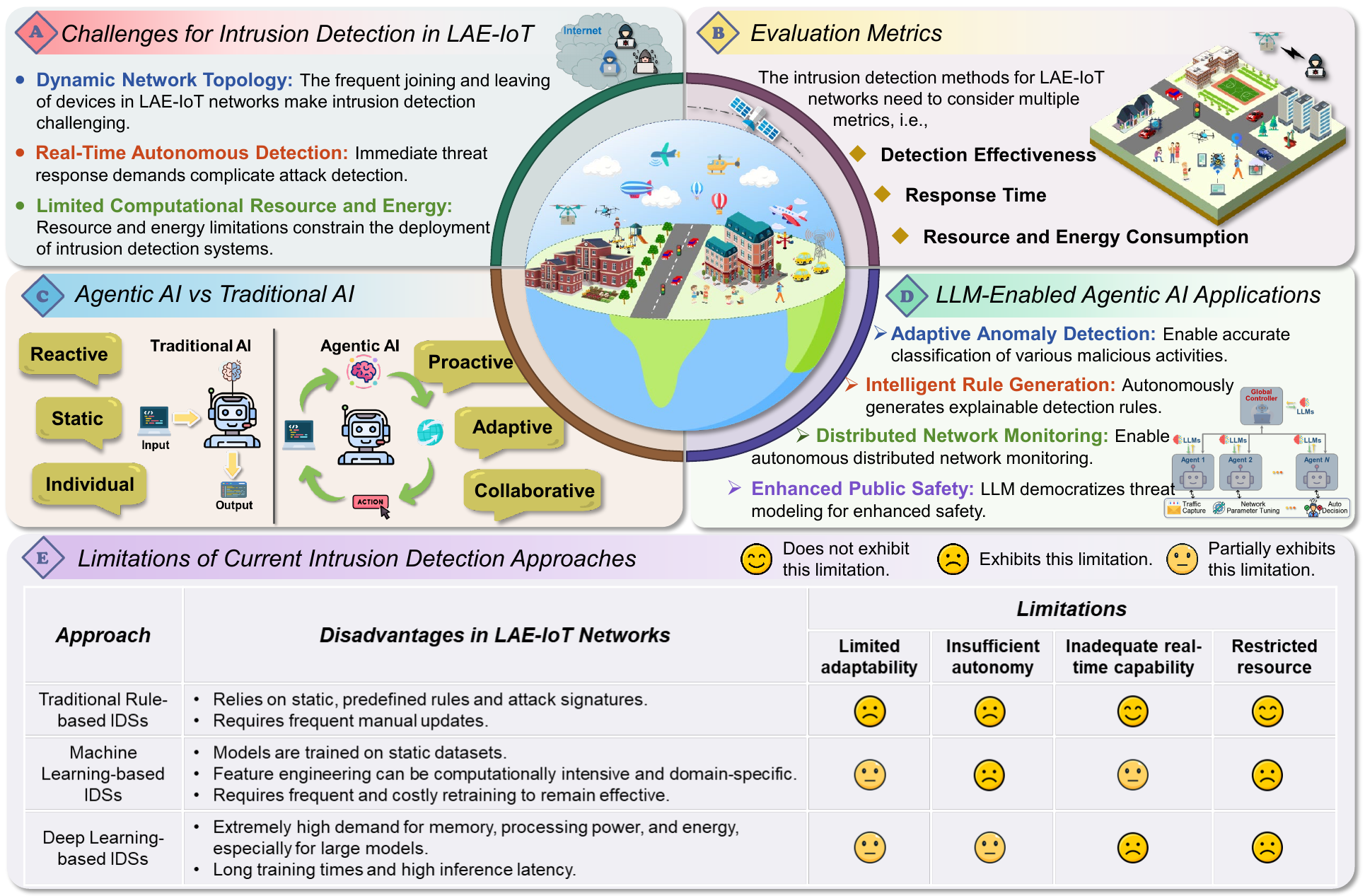}
    \caption{An overview of LLM-enabled agentic LAE-IoT intrusion detection. \emph{Part A} introduces three key challenges for intrusion detection in LAE-IoT networks. \emph{Part B} presents the evaluation metrics of IDSs. \emph{Part C} provides a comparison between agentic AI and traditional AI paradigms. \emph{Part D} gives important LLM-enabled agentic AI applications, including adaptive anomaly detection, intelligent rule generation, distributed network monitoring, and enhanced public safety.}
    \label{fig:Figure_1}
\end{figure*}

\subsection{Challenges for Network Intrusion Detection in LAE-IoT}
\par IDSs have been extensively studied and deployed in traditional network environments, which serve as critical components for monitoring network traffic and identifying malicious activities. Different from terrestrial networks with relatively stable topologies and sufficient computational resources, LAE-IoT networks exhibit unique characteristics, including high mobility, autonomous decision-making requirements, and severe resource constraints. These characteristics make LAE-IoT networks particularly vulnerable to diverse security threats. For example, distributed denial-of-service (DDoS) attacks in LAE-IoT networks can be launched by compromised aerial devices that flood target nodes with overwhelming traffic across multiple altitudes and geographic locations, making source identification challenging. Furthermore, attackers can secretly deploy rogue aerial devices into the network to impersonate legitimate nodes, or launch GPS spoofing attacks by transmitting false positioning signals to mislead UAVs about their locations. Considering these security threats and the unique operational characteristics of LAE-IoT networks, IDSs face the following key challenges, as depicted in Fig.~\ref{fig:Figure_1}~\emph{Part A}.

\subsubsection{Dynamic Network Topology}
\par LAE-IoT networks experience frequent topology changes where devices continuously join and leave the network due to mission requirements and battery limitations. In this case, the high mobility of network nodes makes it challenging to maintain consistent device profiles and establish reliable trust relationships. The authors in~\cite{Ceviz2025} emphasize that the rapid movement patterns of devices in LAE networks create significant obstacles for maintaining stable communication among IDSs. Frequent network reconfiguration creates a scenario where traditional security baselines become obsolete almost immediately after they are established, as the reference points for normal network behavior continuously shift with each topology change.
\par On the other hand, malicious entities can exploit the inherent network instability to mask their presence and activities in specific LAE scenarios. For instance, in aerial-ground communications, adversaries can exploit handoff procedures between aerial base stations and terrestrial gateways to launch man-in-the-middle attacks. Moreover, attackers may strategically choose to launch attacks during LAE-IoT network re-clustering events, where devices dynamically reorganize into new groups, which makes it extremely difficult to distinguish between authorized topology updates and malicious node injection~\cite{Ceviz2025}. Additionally, in multi-tier LAE-IoT architectures, including high-altitude platforms, low-altitude UAVs, and ground IoT devices, malicious actors can exploit the vertical handover processes between different altitude layers to establish unauthorized connections before detection mechanisms can adapt to the new 3D topology~\cite{Mmedhi2025, Hadi2024}.

\subsubsection{Real-Time Detection Requirements}
\par LAE-IoT networks demand immediate threat identification and response capabilities to ensure mission continuity and operational safety in dynamic environments. Specifically, IDSs in LAE-IoT environments need to capture network traffic in real-time and identify security threats within seconds to prevent mission disruption or safety hazards. The authors in~\cite{Birnbaum2015} illustrate the complexity of real-time operational monitoring in aerial environments, where continuous flight data needs to be processed instantaneously to detect anomalies in real-time during active operations, but the volume of this data creates substantial processing burdens. Moreover, LAE-IoT networks require autonomous detection systems that can operate independently without human intervention across vast areas. In this case, IDSs need to make accurate threat assessment decisions independently while coordinating with distributed agents to maintain network-wide security coverage without external guidance.

\subsubsection{Limited Computational Resources and Energy}
\par LAE-IoT devices operate under severe computational resource limitations that constrain the deployment of complex intrusion detection mechanisms. Conventional IDSs typically require substantial computational overhead for complex pattern recognition and threat analysis. However, the miniaturized hardware platforms commonly used in aerial IoT devices possess limited processing power, memory capacity, and storage resources that are insufficient for these methods. Recent studies, such as~\cite{Hadi2024} demonstrate that advanced deep learning-based detection models can achieve high accuracy rates exceeding 90\%, but these models demand significant computational resources that far exceed the capabilities of resource-constrained aerial devices. Meanwhile, energy constraints present equally critical challenges for IDSs in LAE-IoT networks, as battery-powered aerial devices must carefully manage power consumption to maintain operational functionality. Energy consumption directly impacts device performance and reliability, yet traditional intrusion detection approaches often ignore energy efficiency considerations in their design and implementation.

\subsection{Evaluation Metrics for LAE-IoT Intrusion Detection}
\par The evaluation of IDSs in LAE-IoT networks requires metrics that tackle unique operational challenges. In contrast to conventional networks with stable communication and abundant resources, LAE-IoT networks operate under stringent constraints, including dynamic topologies, limited processing power, and restricted battery capacity. Thus, evaluation metrics are necessary to balance security effectiveness with operational sustainability, with the details in the following.

\subsubsection{Detection Effectiveness}
\par In LAE-IoT networks, important metrics such as accuracy, precision, and recall are used for evaluating the effectiveness of intrusion detection methods. Specifically, accuracy quantifies the overall correctness of classification decisions, while precision evaluates the reliability of methods by measuring the proportion of detected intrusions that are genuinely malicious. Similarly, recall also focuses on assessing the ability of IDSs to capture actual security breaches. The F1-score provides a balanced assessment by harmonizing precision and recall, which proves especially valuable when dealing with imbalanced datasets common in LAE-IoT scenarios.

\subsubsection{Response Time}
\par Response time constitutes a critical performance metric for IDSs in LAE-IoT networks, which represents the time interval between threat occurrence and system detection. In~\cite{Hadi2024}, the authors note that longer detection time can decrease the practical value of IDSs regardless of their accuracy levels. When IDSs exhibit prolonged response time exceeding several minutes, they provide attackers with sufficient operational windows to execute malicious activities, including hole attacks, sensitive data extraction, and harmful packet forwarding. Consequently, IDSs in LAE-IoT networks need to provide immediate defensive responses after identifying threats.

\subsubsection{Resource and Energy Consumption}
\par Resource utilization and energy consumption are essential performance metrics for resource-constrained and battery-powered LAE-IoT devices, which directly impact operational duration and mission capabilities. Complex event processing-based IDS architectures for IoT environments have been developed with a focus on detection time, memory consumption, and CPU utilization. Similarly, intrusion detection frameworks for energy-constrained IoT devices that leverage collaboration between host devices and edge nodes have been proposed to minimize memory and energy consumption. The authors in~\cite{Arshad2020} design an intrusion detection framework for energy-constrained IoT devices that leverages collaboration between host devices and edge nodes to minimize memory and energy consumption. The framework implements lightweight signature-based detection at individual IoT devices to monitor network traffic and identify known attack patterns, while delegating computationally intensive anomaly-based detection and decision tasks to resource-rich edge routers. This architectural design achieves minimal performance overhead, with device-level components consuming only 274 bytes of ROM and 368-728 bytes of RAM depending on alert aggregation strategies, while edge router CPU power consumption increases negligibly regardless of alert processing load.

\subsection{Limitations of Current Intrusion Detection Approaches}

\par Despite significant advances in intrusion detection technologies, existing approaches exhibit critical limitations when deployed in LAE-IoT environments, as they are primarily designed for traditional network infrastructures, and the details are as follows:

\begin{itemize}
    \item{\textbf{Limited Adaptability:}} Traditional machine learning and deep learning-based IDSs rely on static models that cannot adapt to the rapidly evolving topology and attack patterns of LAE-IoT networks without extensive retraining~\cite{Turukmane2024}.
    
    \item{\textbf{Insufficient Autonomy:}} Deep learning-based IDSs exhibit high inference latency, computational overhead, and real-time deployment issues, while traditional approaches require continuous human intervention for model updates and threshold tuning, which makes them unsuitable for autonomous operation across vast coverage areas with intermittent connectivity.
    
    \item{\textbf{Resource Constraints:}} State-of-the-art detection models demand computational resources and energy consumption levels that far exceed the capabilities of resource-constrained, battery-powered LAE-IoT devices~\cite{Mmedhi2025, Arshad2020}.
\end{itemize}

\par These limitations motivate us to propose novel detection paradigms that can tackle the unique operational constraints of LAE-IoT networks, which we address through LLM-enabled agentic approaches in subsequent sections.

\section{LLM-Enabled Agentic LAE-IoT Intrusion Detection}

\par In this section, we present the characteristics of LLMs and agentic AI. Moreover, we compare agentic AI with traditional AI paradigms and introduce key LLM-enabled agentic AI applications.

\subsection{Overview and Characteristics of LLM and Agentic AI}
\par LLMs are multi-layer deep neural architectures with billions of parameters that provide strong cognitive reasoning capabilities through extensive pre-training on vast amounts of data~\cite{Gu2025, Sapkota2026}. In LAE-IoT security, LLMs excel at understanding cybersecurity patterns, threat taxonomies, and network behavior analysis through their extensive pre-trained knowledge bases and semantic reasoning capabilities.
\par Agentic AI establishes a robust cognitive architecture for intrusion detection through four interconnected components, i.e., perception, memory, reasoning, and action. Different from LLMs that process individual queries independently, agentic AI systems maintain persistent threat memory and environmental awareness to pursue long-term security objectives autonomously in LAE-IoT networks. Specifically, the perception component captures and processes multimodal data, which subsequently feeds into the memory module that can utilize retrieval-augmented generation (RAG) mechanisms to retrieve and retain historical knowledge. The reasoning component, often powered by LLMs serving as cognitive engines, may employ sophisticated reasoning methods, such as chain-of-thought, to examine retrieved contextual data and design multi-step strategies. The action component implements strategic decisions through synchronized policy deployment and collaborative multi-agent coordination mechanisms. This agentification workflow facilitates continuous enhancement via feedback mechanisms, thus enabling agents to evolve their threat detection and response capabilities in LAE-IoT environments.

\subsection{Comparison of Agentic AI and Traditional AI Paradigms}
\par The shift from traditional AI to agentic AI offers significant advantages for LAE-IoT intrusion detection.

\subsubsection{Reactive and Proactive Response}
\par Traditional IDSs operate primarily in reactive modes and respond to attack events only after detection. In contrast, agentic AI exhibits autonomy by decomposing high-level objectives into executable sub-objectives, planning multi-step response strategies, and initiating actions based on reasoning about potential future threats~\cite{Zhang2025}. For instance, an agent can leverage LLMs to interpret ambiguous or incomplete logs as anomalous patterns indicative of a potential attack, thus enabling it to invoke network management tools to reconfigure UAV swarm defenses.

\subsubsection{Static Learning and Adaptive Intelligence}
\par Traditional machine learning-based IDSs rely on static models that require offline retraining to handle new threats. Agentic AI achieves adaptive intelligence through an integration of contextual memory within its reasoning processes~\cite{Zhang2025, Sapkota2026}. For instance, when encountering novel GPS spoofing techniques, the system can analyze attack characteristics against stored threat patterns and immediately modify its trajectory validation algorithms based on learned similarities.

\subsubsection{Individual Detection and Collaborative Intelligence}
\par Traditional IDSs operate as independent processing units that make decisions based solely on local observations. Agentic AI enables collaborative intelligence through orchestrated interaction among multiple agents~\cite{Sapkota2026}. For example, agents on different nodes in LAE-IoT networks can share local threat data, and the central controller analyzes reports to identify coordinated attacks and allocate tasks based on node capabilities and real-time environmental conditions.

\subsection{LLM-Enabled Agentic AI Applications}
\subsubsection{Adaptive Anomaly Detection}
\par The emergence of robust and adaptive anomaly detection has become increasingly critical as cybersecurity threats continue to evolve in sophistication and scale. Traditional IDSs, while effective in controlled environments, often struggle with adaptability when faced with novel attack vectors or dynamic operational contexts. The authors in~\cite{Honnalli2025} design an anomaly detection agent designed for electric vehicle charging networks that integrates GPT-4 series models with RAG technology. For LAE-IoT intrusion detection, this approach can tackle the challenge that UAVs exhibit diverse operational profiles. In this case, the RAG mechanism can be adapted with domain-specific knowledge, such as communication protocols, standard flight parameters, and known aerial threat signatures, to perform context-aware anomaly detection that is resilient to the dynamic topology in LAE-IoT networks. Compared to approaches that depend on historical anomaly patterns and cannot integrate real-time contextual knowledge, the system proposed in~\cite{Honnalli2025} accurately classifies attacks without requiring predefined rule sets or extensive retraining, which effectively defends against ever-evolving cyber threats and achieves superior performance compared to traditional LSTM-based methods in detecting complex cyber threats.

\subsubsection{Intelligent Rule Generation}
\par Compared to closed-box machine learning-based IDSs, the rule-based IDSs offer superior interpretability, deterministic behavior, and real-time performance. Traditional rule generation methods face significant scalability challenges, as they rely heavily on manual expertise and domain knowledge, resulting in lengthy development cycles, time-consuming updates and maintenance, and limited adaptability to emerging attack vectors. Motivated by the generalization and autonomy of LLM and agentic AI, the authors in~\cite{Gu2025} introduce a framework where a novel agentic AI architecture where a detection agent proposes anomaly detection rules, a repair agent corrects syntax errors, and a review agent validates rule accuracy through iterative feedback loops. This collaborative agentic framework is advantageous for LAE-IoT networks to overcome the key challenge that conventional deep learning models are often too resource-intensive for deployment on UAVs. By offloading computation-intensive tasks of multi-agent collaboration and LLM-based reasoning to ground-based servers, the framework autonomously generates detection rules that adapt to different anomaly patterns. Then, these rules can be deployed onto resource-constrained aerial nodes, enabling real-time, low-latency anomaly detection during flight missions.

\subsubsection{Distributed Network Monitoring}
\par LAE-IoT networks, characterized by dynamic UAV networks, edge computing clusters, and distributed sensor arrays, present unique challenges for network security monitoring due to their inherently decentralized nature and resource constraints. For tackling the scalability and responsiveness challenges in distributed network monitoring, the authors in~\cite{Zambare2025} propose a dual-tier architecture with lightweight micro-agents on individual nodes for local analysis and a central controller for system-wide threat correlation. Inspired by this study, this dual-tier architecture can be adapted for LAE-IoT networks, where UAVs and other aerial nodes are deployed with lightweight micro-agents to perform local traffic analysis and preliminary anomaly detection. Simultaneously, a ground station or a designated leader UAV can function as the centralized controller for aggregating intelligence from these distributed agents to identify coordinated attacks.

\subsubsection{Enhanced Public Safety}
\par As the LAE rapidly expands with increasing deployment of IoT devices, urban air mobility systems, and autonomous aerial vehicles for logistics and transportation services, public safety has become a critical concern, where malicious intrusions can directly threaten civilian safety, disrupt emergency services, and compromise critical infrastructure operations. Recent study~\cite{Zisad2025}  introduces an agentic AI system that leverages few-shot learning and the Gemini model for democratizing threat modeling processes. Compared to manual threat model generation requiring over 40 hours, the system generates threat models in approximately 30 seconds. In the LAE-IoT networks, this system identifies specific vulnerabilities and potential attacks, which in turn provides timely and relevant intelligence to the intrusion detection framework, thus guiding it on what malicious behaviors and anomalies to prioritize for detection.

\subsection{Lessons Learned}
\par Building on prior research, LLM-enhanced agentic AI has demonstrated significant potential in intrusion detection for LAE-IoT networks. The cognitive reasoning capabilities of LLMs and the proactive, adaptive, and collaborative nature of agentic AI can tackle the limitations of traditional IDSs. However, current studies often rely on individual agent capabilities or centralized detection approaches. Although some studies consider using multiple agents, they lack a general framework that integrates perception, memory, reasoning, and action components for intrusion detection in dynamic LAE-IoT environments.
\begin{figure*}
    \centering
    \includegraphics[width=0.9 \linewidth]{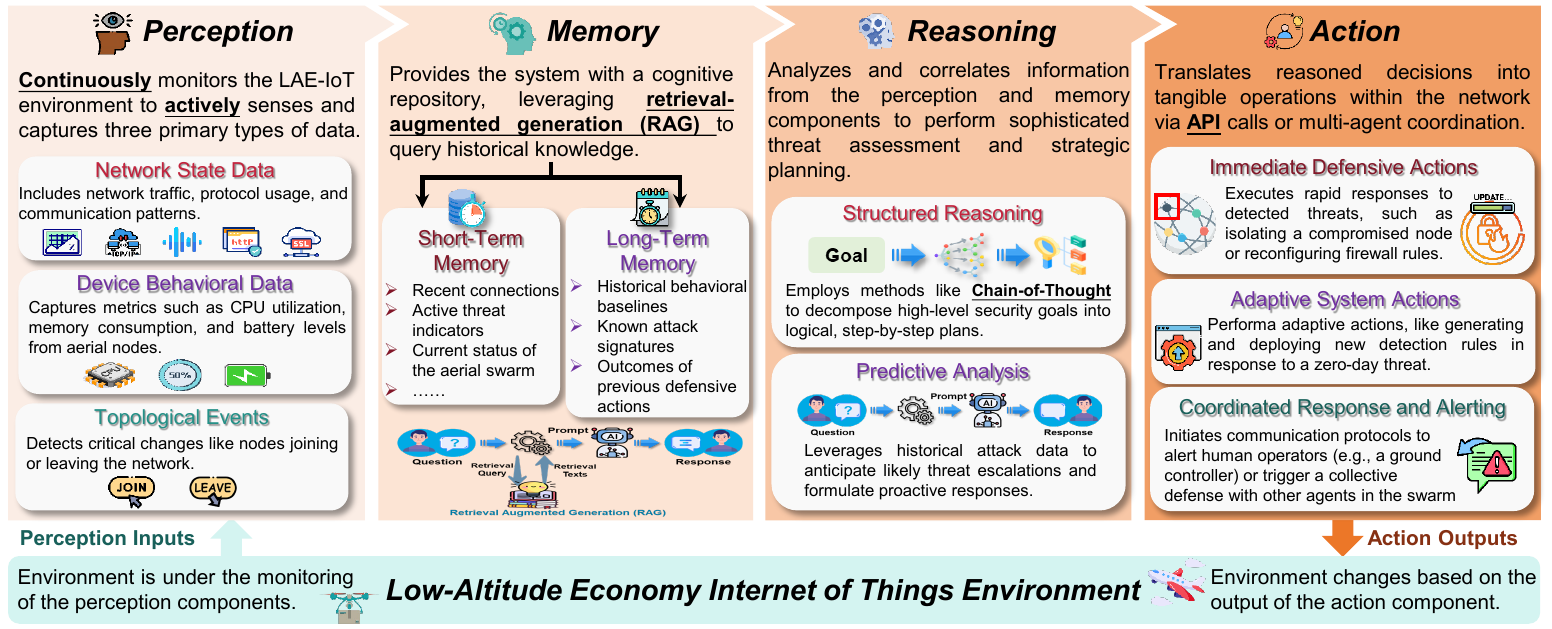}
    \caption{The general agentic AI framework for intrusion detection in LAE-IoT networks. The framework operates in a continuous loop, consisting of four core components: perception, memory, reasoning, and action, which interact with the LAE-IoT environment.}
    \label{fig:Figure_2}
\end{figure*}

\section{Agentic AI-based Intrusion Detection for LAE-IoT Networks}
\par Building upon the analysis of LLM-enabled agentic AI applications, we now present a general framework design that addresses the specific challenges identified in LAE-IoT intrusion detection.

\begin{itemize}
    \item{\textbf{Perception Component:}} The perception module serves as the sensory interface of the agentic system for monitoring the LAE-IoT environment. Specifically, this component continuously inspects network traffic for signatures of known attacks, e.g., DDoS and port scanning, and anomalous communication patterns. Moreover, it also captures device-level behavioral data, such as unusual CPU spikes or memory usage, which could indicate a compromised node. Furthermore, it tracks topological events, such as an unscheduled device joining the swarm, which could signal a rogue node injection attempt.

    \item{\textbf{Memory Component:}} This component serves as the knowledge base of agentic systems, which provides the context needed for accurate intrusion detection. The short-term memory maintains immediate operational data, including recent network connections and the current device status across the aerial swarm. The long-term knowledge repository stores baselines of normal network behavior, features of known LAE-IoT attacks, and a history of past incidents and their response outcomes. By leveraging RAG mechanisms, the system can query this repository to retrieve contextually relevant information, recognize emerging threats, and refine its defensive strategies over time, thus overcoming the static nature of traditional IDSs.

    \item{\textbf{Reasoning Component:}} Serving as the cognitive engine, often powered by LLMs, this module analyzes and correlates information from the perception and memory components to perform sophisticated threat assessments. It performs tasks crucial for intrusion detection, such as correlating low-confidence alerts from multiple UAVs to uncover a distributed attack, distinguishing false positives from genuine threats to conserve resources, and formulating multi-step response plans.

    \item{\textbf{Action Component:}} This component executes the defensive countermeasures decided by the reasoning component. These operations can be executed through direct API calls to network functions or via multi-agent coordination protocols to orchestrate a collective defense.
\end{itemize}

\par As shown in Fig.~\ref{fig:Figure_2}, these four components operate in a continuous perception-reasoning-action loop for autonomous intrusion detection. For instance, when perceiving an unrecognized device attempting to join the swarm with unusual communication patterns, the agent queries memory to confirm this unscheduled event and cross-references against known rogue node injection signatures. Then, the reasoning component correlates this information, assesses it as a high-probability intrusion, and formulates a multi-step response plan. Finally, the action module executes the plan, i.e., by isolating the suspicious node and alerting a ground controller, while the perception module continues monitoring, thus feeding new information back into the loop for ongoing assessment and adaptation. This iterative cycle empowers the agent to be proactive, context-aware, and self-improving, thus providing effective intrusion detection for LAE-IoT networks.

\begin{figure*}
    \centering
    \includegraphics[width=0.99 \linewidth]{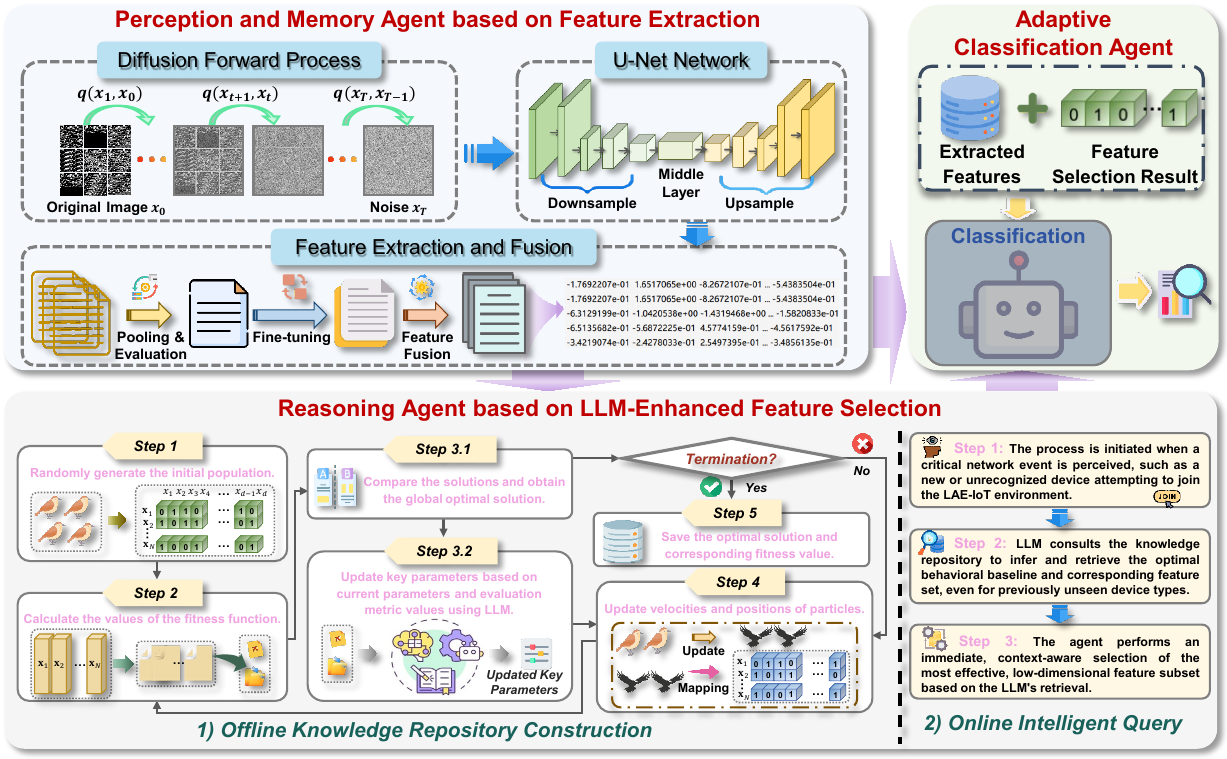}
    \caption{The multi-agent collaborative intrusion detection framework, which consists of three agents, namely, a perception and memory agent based on feature extraction for network traffic processing, a reasoning agent based on LLM-enhanced feature selection for dimensionality reduction, and a classification agent for threat identification.}
    \label{fig:Figure_3}
\end{figure*}

\section{Case Study: LLM-Enabled Agentic Intrusion Detection Framework for LAE-IoT Networks}
\par In this section, we propose an LLM-enabled agentic AI framework to enhance the effectiveness of intrusion detection for LAE-IoT networks.
\subsection{LLM-Enabled Agentic Intrusion Detection Framework}
\par  As shown in Fig.~\ref{fig:Figure_3}, we propose an LLM-enhanced agentic intrusion detection framework for LAE-IoT networks that operates within the continuous perception-reasoning-action loops. The details are as follows.

\par \textbf{Perception and Memory Agent based on Feature Extraction:} This agent embodies the perception and memory functions of the framework, which is responsible for actively perceiving the underlying data distributions to build a stable, generalizable knowledge base. To leverage powerful representation learning techniques from computer vision and avoid manual feature engineering, the agent first transforms raw traffic sessions into two-dimensional grayscale images. Then, the agent employs a self-supervised denoising diffusion probabilistic model (DDPM) on these visual representations for learning the underlying data distribution of benign traffic images through the generative pre-training rather than memorizing instance-specific network flows. The learned universal features from this self-supervised process form the memory mechanism $\mathcal{M}$ of the agentic framework. This memory comprises universal representations $\mathcal{M} = \{\phi_1, \phi_2, \ldots, \phi_n\}$ that persist across varying network topologies and traffic patterns, which serve as prior knowledge for rapid adaptation to emerging threats. This agent effectively decouples the learned features from the high-level network configuration. While the network graph evolves as nodes join or leave, the underlying structural representation of benign traffic remains largely consistent. In contrast, malicious traffic typically manifests as visually distinctive, high-entropy patterns or noisy textures. Thus, the effectiveness of this agent persists without requiring constant retraining for every topological shift. It generalizes the visual characteristics of normal versus anomalous behavior, and this generalized knowledge stored in memory $\mathcal{M}$ provides a stable foundation for detection in highly dynamic aerial environments, thereby enabling few-shot learning capabilities when new attack types emerge.

\par \textbf{Reasoning Agent based on LLM-Enhanced Feature Selection:} This agent embodies the reasoning capability of the framework, which leverages LLMs to perform intelligent decision-making and knowledge-driven optimization. This agent serves as the cognitive core of the framework, operating in a two-stage process, i.e., an offline LLM-guided knowledge generation phase and an online real-time intelligent query phase. This design leverages the deep reasoning of LLMs for complex, offline optimization while enabling lightweight, rapid execution on resource-constrained aerial devices.
\par \textit{1) Offline Knowledge Repository Construction:} Before deployment, the framework executes an intensive optimization process on a ground server to build a comprehensive knowledge repository. Specifically, we employ the LLM as an autonomous algorithm expert to supervise and dynamically tune the feature selection process. The LLM observes the iterative progress of particle swarm optimization (PSO) and adaptively adjusts its key parameters. Through advanced reasoning processes, the LLM analyzes a summary state of the population after each optimization epoch, including fitness value and population diversity, to diagnose performance issues, such as premature convergence or search stagnation. Leveraging its extensive built-in knowledge of swarm intelligence principles, it then reasons about this diagnosis to formulate a corrective strategy, for instance, concluding that the swarm is trapped in a local optimum and requires greater exploration. This cognitive process culminates in the dynamic generation of a new set of key parameters for PSO, e.g., inertia weight, cognitive, and social coefficients, which are then injected into the subsequent epoch to effectively steer the search towards a more optimal and robust solution.
\par The outcome of this LLM-guided process is a pre-compiled knowledge repository that maps device profiles to feature subsets discovered through a truly intelligent and robust search.
\par \textit{2) Online Intelligent Query:} During real-time operation, when a new device joins the network, the framework perceives this event. Then, leveraging its reasoning capabilities, the LLM formulates an intelligent query to the pre-compiled knowledge repository, in which the LLM can infer the most appropriate entry even for novel or slightly varied device types, demonstrating zero-shot generalization. This ensures an immediate, context-aware selection of the most effective, low-dimensional feature subset, which significantly reduces the computational load for the subsequent classification agent and enables real-time, efficient threat detection.

\par \textbf{Adaptive Classification Agent:} This agent serves as the decisive action component of the framework, executing the final traffic classification. Its operation is initiated by perceiving the real-time resource status of the device, including battery level and CPU load. Based on the real-time resource status of devices, the agent queries a dedicated memory module that stores a pool of pre-trained classification models. This pool includes models of varying complexities, from highly efficient options, such as LightGBM to more resource-intensive, higher-accuracy alternatives. Subsequently, it applies this chosen model to the reduced feature vector provided by the reasoning agent to perform its primary action, i.e., classifying the traffic as benign or malicious. 
\par To improve the efficiency of this framework, we propose a lightweight deployment strategy. The computationally intensive processes, such as the training of the DDPM for feature extraction and the LLM-driven optimization of feature selection configurations, are performed offline on ground-based servers or powerful edge nodes. The resulting optimized models and feature sets are deployed to the aerial devices. In this case, the LAE-IoT nodes are only required to execute the highly efficient online tasks, i.e., applying the pre-selected feature mask and running inference with the lightweight classifier. This offline-online architecture ensures that real-time threat detection is performed with minimal latency and resource consumption.

\subsection{Numerical Results}
\par In our performance evaluation, we consider a scenario where UAVs serve as mobile data collectors for a distributed ground-based IoT network. In this scenario, compromised ground IoT devices can be leveraged to launch internal network intrusions. To validate the effectiveness of our framework against such threats, we specifically utilize the Edge-IIoTset\footnote{https://ieee-dataport.org/documents/edge-iiotset-new-comprehensive-realistic-cyber-security-dataset-iot-and-iiot-applications}, which contains realistic attack patterns originating from IoT devices, such as DDoS, port scanning, and SQL Injection. We supplement this with two other general traffic datasets, USTC-TFC\footnote{https://github.com/davidyslu/USTC-TFC2016} and ISCX-VPN\footnote{https://www.unb.ca/cic/datasets/vpn.html}, to ensure comprehensive evaluation. We compare the proposed framework against several state-of-the-art baseline methods, including supervised learning approaches, \emph{i.e.}, 2D-CNN and RBLJAN, and self-supervised learning methods, \emph{i.e.}, YaTC and MTC-MAE~\cite{Xu2024, Yang2025}.

\par To demonstrate the autonomous and proactive capabilities in the framework, we introduce detection latency as a key metric to measure the ability of the framework to rapidly adapt to new or unknown attacks with minimal human intervention. Detection latency is defined as the total time from encountering a new threat to achieving effective detection, which is composed of the expert labeling time ($T_{\text{labeling}}$) and model training time ($T_{\text{training}}$). Since $T_{\text{labeling}}$ is typically much greater than $T_{\text{training}}$, detection latency is predominantly determined by the amount of labeled data required to achieve a target performance. In this case, a lower requirement for labeled data directly translates to faster adaptation and greater autonomy. Moreover, we employ four conventional classification metrics to assess detection accuracy, which are accuracy, precision, recall, and macro F1-score.

\begin{figure*}
    \centering
    \includegraphics[width=0.99 \linewidth]{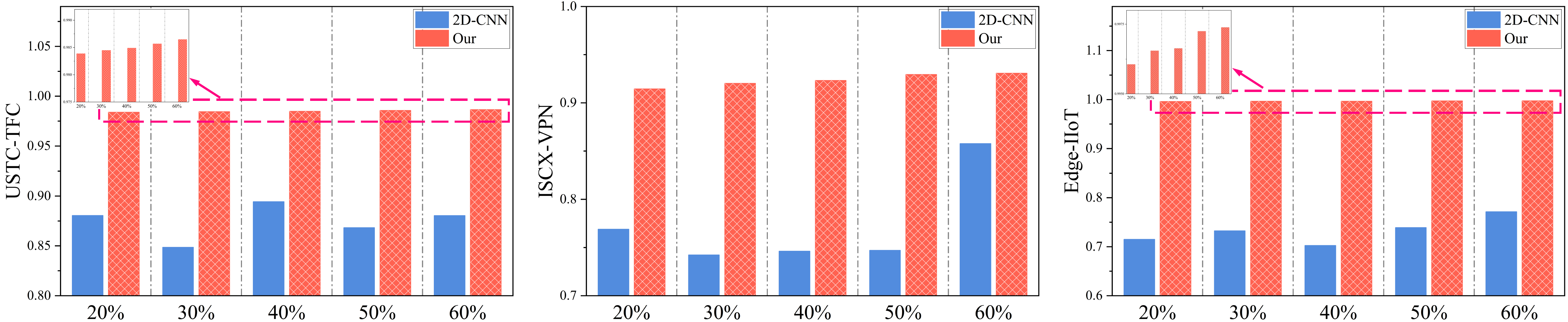}
    \caption{Classification accuracy comparison between the proposed framework and 2D-CNN baseline across varying percentages of labeled data on three datasets.}
    \label{fig:finetune_results}
\end{figure*}
\par \textit{1) Detection Latency Analysis:} Fig.~\ref{fig:finetune_results} shows the comparison of classification accuracy between the proposed framework and the 2D-CNN baseline across varying amounts of labeled data on three benchmark datasets. As can be seen, the proposed framework surpasses the peak accuracy of the baseline using significantly less labeled data across all three datasets, which indicates superior label efficiency of our framework. Since annotation time dominates the deployment timeline for emerging threat detection, the reduction in labeling requirements means our framework can be deployed faster than traditional supervised approaches, thus enabling rapid response to evolving threats in dynamic LAE-IoT environments. The reason may be that the features stored in memory $\mathcal{M}$ through self-supervised pre-training enable rapid few-shot adaptation. Moreover, across all three datasets spanning different network characteristics, the proposed framework maintains high accuracy regardless of the proportion of labeled data used. This superior and stable performance highlights the robustness of the learned representations in memory $\mathcal{M}$, which serve as reliable prior knowledge that enables effective threat detection even in data-scarce scenarios. These results demonstrate that the LLM-enhanced agentic framework can effectively detect attacks with minimal human intervention and significantly accelerate deployment cycles for emerging threats, which tackles the critical challenge of rapid autonomous detection in dynamic aerial environments where traditional supervisory methods would introduce unacceptable delays.

\begin{figure*}
    \centering
    \includegraphics[width=0.99 \linewidth]{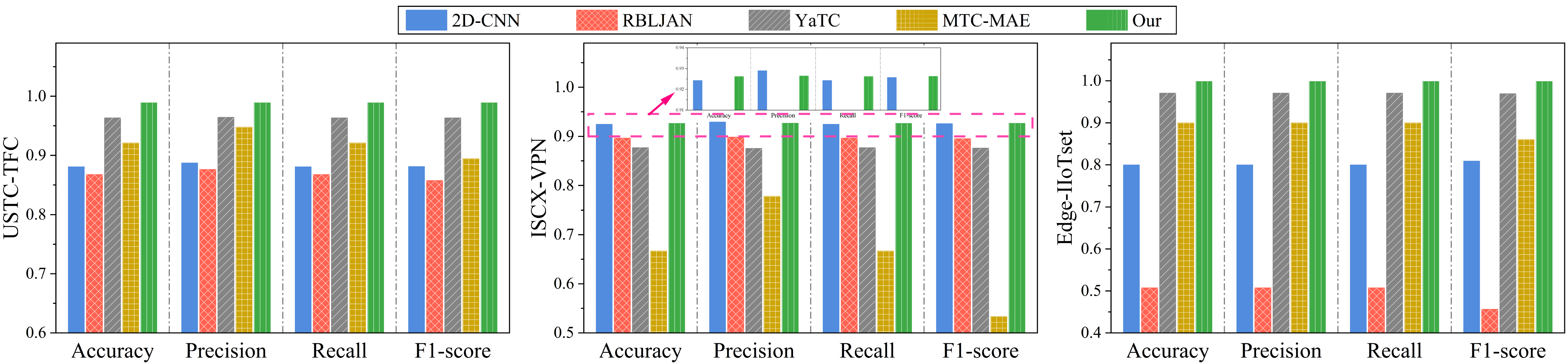}
    \caption{Performance comparison between our proposed framework and baseline methods on three datasets.}
    \label{fig:results}
\end{figure*}

\par \textit{2) Classification Performance Analysis:} Fig.~\ref{fig:results} presents the performance comparison results across the three datasets. As can be seen, the proposed framework outperforms baseline methods and achieves an overall classification accuracy of over 90\% on three datasets. Based on these results, we can draw several key conclusions.
\par \emph{First}, the superior performance of our LLM-enhanced agentic framework compared to traditional supervised approaches, \emph{i.e.}, 2D-CNN, RBLJAN, demonstrates the effectiveness of integrating cognitive reasoning capabilities with distributed agent coordination for LAE-IoT intrusion detection.
\par \emph{Second}, although advanced self-supervised methods, \emph{i.e.}, YaTC and MTC-MAE, show competitive performance through representation learning, their performance still falls short compared to our agentic framework, which indicates the superior ability of our multi-agent architecture to capture complex threat patterns through collaborative intelligence and adaptive feature optimization. The LLM-guided feature selection agent effectively eliminates redundant information while preserving the most discriminative characteristics crucial for intrusion detection in dynamic LAE-IoT networks. 
\par \emph{Third}, the consistent performance of our framework across different datasets demonstrates its strong generalization capability and adaptability to diverse traffic patterns encountered in LAE applications, making it particularly valuable for securing heterogeneous aerial IoT environments with varying device types and operational contexts.
\par \emph{Fourth}, the balanced precision and recall metrics of our framework further demonstrate its ability to minimize both false positives and false negatives, which is crucial for LAE-IoT security systems where false alarms can drain limited battery resources and missed threats can compromise mission-critical operations. This balance is achieved through the intelligent coordination between our perception and memory agent based on feature extraction, reasoning agent based on LLM-enhanced feature selection, and adaptive classification agent.

\section{Open Challenges and Future Directions}
\par In this section, we present future research directions and discuss several open challenges that need to be overcome for practical deployment of LLM-enhanced agentic intrusion detection in LAE-IoT networks.

\par \textbf{(1) Data Scarcity and Heterogeneous Data Fusion:} A pressing challenge is the lack of comprehensive, standardized datasets that accurately represent the unique traffic patterns and attack scenarios of LAE-IoT. Future work should consider integrating heterogeneous data sources beyond network traffic. By enabling agentic frameworks to process and correlate diverse data streams, including flight telemetry and device behavioral logs, we can establish deeper context, which allows for the detection of coordinated attacks, such as a minor network intrusion coinciding with an unexpected deviation in flight path.
\par \textbf{(2) Robust Multi-Agent Coordination and Orchestration:} LAE-IoT networks frequently experience communication disruptions due to interference, weather, or physical obstacles. Our current framework assumes reliable inter-agent communication, but real-world deployments demand greater resilience. Future research should investigate hierarchical agent orchestration. In this model, local agents on UAVs could make autonomous decisions during communication outages, while higher-level agents on ground stations or edge nodes coordinate swarm-level intelligence when connectivity is restored. Techniques such as hierarchical reinforcement learning and distributed LLM inference can ensure efficient, scalable security coverage while minimizing communication overhead.
\par \textbf{(3) Lightweight Edge-Optimized LLM Integration:} Energy efficiency and computational constraints are critical limitations in LAE-IoT networks, where aerial devices often operate with limited battery power and processing capabilities. Future research should explore techniques, such as model compression, knowledge distillation, and adaptive inference scheduling, that can further optimize the trade-off between detection accuracy and resource consumption.

\section{Conclusion}
\label{sec:conclusion}
\par In this article, we have explored the application of LLM-enhanced agentic AI techniques to improve the performance of intrusion detection in LAE-IoT networks. We have analyzed the unique challenges for LAE-IoT intrusion detection, including dynamic network topologies, real-time autonomous detection demands, and resource constraints. Subsequently, we have proposed an innovative multi-agent collaborative framework that leverages the cognitive reasoning capabilities of LLMs to guide autonomous security behaviors through three agents, \emph{i.e.}, feature extraction, intelligent feature selection, and adaptive classification. Through experimental validation on three diverse datasets, our proposed framework has demonstrated exceptional performance. These findings underscore the critical role of agentic AI in enhancing LAE-IoT security and highlight the need for further exploration of collaborative intelligence architectures for securing the emerging LAE ecosystem.

\bibliographystyle{IEEEtran}
\bibliography{magazine_ref}

\end{document}